\documentclass[conference]{IEEEtran}
\IEEEoverridecommandlockouts
\usepackage{cite}
\usepackage{amsmath,amssymb,amsfonts}
\usepackage{amsmath, bm}
\usepackage{algorithmic}
\usepackage{algorithm}
\usepackage{graphicx}
\usepackage{textcomp}
\usepackage{xcolor}
\usepackage{color}
\usepackage{array}
\usepackage{ulem}
\def\BibTeX{{\rm B\kern-.05em{\sc i\kern-.025em b}\kern-.08em
		T\kern-.1667em\lower.7ex\hbox{E}\kern-.125emX}}

\begin{document}
\title{MIMO Precoding Design with QoS and Per-Antenna Power Constraints
}

\author{\IEEEauthorblockN{Kaiyi Chi, Yingzhi Huang, Qianqian Yang, Zhaohui Yang, Zhaoyang Zhang}
    \IEEEauthorblockA{
        College of Information Science and Electronic Engineering, Zhejiang University, Hangzhou 310027, China \\
        E-mail: \{kaiyichi, yingzhihuang, qianqianyang20, yang\_zhaohui, ning\_ming\}@zju.edu.cn}
    
}

\maketitle
\begin{abstract}
    Precoding design for the downlink of multiuser multiple-input multiple-output (MU-MIMO) systems is a fundamental problem. In this paper, we aim to maximize the weighted sum rate (WSR) while considering both quality-of-service (QoS) constraints of each user and per-antenna power constraints (PAPCs) in the downlink MU-MIMO system. To solve the problem, we reformulate the original problem to an equivalent problem by using the well-known weighted minimal mean square error (WMMSE) framework, which can be tackled by iteratively solving three subproblems. Since the precoding matrices are coupled among the QoS constraints and PAPCs, we adopt alternating direction method of multipliers (ADMM) to obtain a distributed solution. Simulation results validate the effectiveness of the proposed algorithm.
\end{abstract}


\section{Introduction}
The multiuser multiple-input multiple-output (MU-MIMO) technique has become a key technology for the next generation wireless communication systems due to its high spectral efficiency \cite{mumimo}. To fully exploit the potential of the MU-MIMO system, it is essential to design the precoding algorithms to mitigate the interference among users \cite{radar-qos}. 

In the past decades, researchers have proposed numerous precoding methods to maximize the weighted sum rate (WSR) for the downlink transmission of the MU-MIMO systems. There are usually two different types of classical methods: intuition-based algorithm and optimization-based iterative algorithm. Intuition-based algorithms, such as zero-forcing (ZF) \cite{zf} and regularized ZF (RZF) \cite{rzf}, usually have low complexity, which are preferred in the practical system. However, these algorithms are not aimed to solve the WSR problem directly, thus leading to performance loss. On the contrary, the optimization-based iterative methods are designed to directly solve the WSR problem. A widely used approach is the weighted minimum mean-square error (WMMSE) algorithm proposed in \cite{wmmse}, where the original WSR problem is transformed into an equivalent minimum mean-square error (MMSE) problem and then the MMSE problem is solved by block coordinate descent (BCD) method in an iterative manner. Furthermore, the authors in \cite{rethinking} proposed a modified WMMSE algorithm with low complexity by exploiting the structure of the optimal transmit beamforming vector.   

Recently, with the development of deep learning (DL), numerous DL-based models have been designed to address the communication problems and these models have demonstrated superiority over the traditional methods \cite{dlbased}. In \cite{deepunfolding}, the authors proposed a deep-unfolding network, which applied a neural network to approximate the iterative WMMSE algorithm. Compared with the standard WMMSE, the deep-unfolding-based method can reduce complexity while maintaining performance. Furthermore, the work in \cite{learningwmmse} explored the structure of the optimal precoding, then employed a customized neural network to approximate the mapping from the channels to the precoding matrix, which also reduced complexity significantly. However, compared with traditional algorithms, DL-based algorithms may suffer from generalization issues.

The beamforming design works mentioned above focus on maximizing the WSR of the system without considering any rate requirements, which may lead to unfairness among users. To address the issue, the authors in \cite{qos} investigated a WSR maximization problem with taking the quality-of-service (QoS) constraints of users into consideration. Furthermore, those works consider sum-power constraint (SPC), however, in practice, power amplifier of each transmit antenna has its own power budget, hence it is more practical to consider the per-antenna power constraints (PAPCs) \cite{papc}. The work in \cite{papc_power_min} utilized downlink-uplink duality to minimize the total power transfer with SINR constraints of each user and PAPCs in a multiple-input single-output (MISO) system. 

In this paper, we aim to maximize the WSR of the downlink MU-MIMO system considering both QoS constraints and PAPCs. To solve the problem, we exploit the framework of WMMSE to reformulate the original problem into an equivalent problem which can be solved by BCD algorithm. To deal with the coupled variables among the QoS constraints and PAPCs, we adopt the alternating direction method of multipliers (ADMM) algorithm \cite{admm}, which introduces auxiliary variables to decouple the QoS constraints and PAPCs. Simulation results validate the effectiveness of the proposed algorithm.

The rest of the article is organized as follows. Section \uppercase\expandafter{\romannumeral2} introduces the system model and formulates the problem. Section \uppercase\expandafter{\romannumeral3} presents the proposed solution.
Section \uppercase\expandafter{\romannumeral4} presents the simulation results to demonstrate the effectiveness of the proposed algorithm. Section \uppercase\expandafter{\romannumeral5} concludes the paper.

\section{System Model and Problem Formulation}
We consider a downlink MU-MIMO system, where a base station (BS) equipped with $N_{t}$ transmit antennas serves $K$ users each with $N_{r}$ receive antennas. The received signal $\mathbf{y}_{k}$ at the $k$-th user can be given by
\begin{equation}
    \mathbf{y}_{k}=\mathbf{H}_{k} \mathbf{V}_{k} \mathbf{s}_{k}+\sum_{j=1, j \neq k}^{K} \mathbf{H}_{k} \mathbf{V}_{j} \mathbf{s}_{j}+\mathbf{n}_{k}, \forall k \in \mathcal{K},
\end{equation}
where $\mathbf{H}_{k} \in \mathbb{C}^{N_{r} \times N_{t}}$ denotes the channel matrix between the BS and user $k$, $\mathbf{V}_{k} \in \mathbb{C}^{N_{t} \times d}$ is the precoding matrix that the BS uses to process the transmit signal vector $\mathbf{s}_{k} \in \mathbb{C}^{d \times 1}$ for user $k$, $d$ is the number of data streams, the transmit signal vector satisfies $\mathbb{E}\left[\mathbf{s}_{k} \mathbf{s}_{k}^{H}\right]=\mathbf{I}$, and it is assumed that the transmitted data streams among different users are independent, $\mathbf{n}_{k} \in \mathbb{C}^{N_{r} \times 1}$ denotes the additive white Gaussian noise with the distribution $\mathcal{C N}\left(0, \sigma^{2} \mathbf{I}\right)$, and $\mathcal{K}=\{1,2, \ldots, K\}$ denotes the set of users. 

We aim at optimizing the precoding matrices to maximize the weighted sum rate of the system subject to both QoS constraints and PAPCs. Thus, the problem is formulated as
\begin{subequations}
    \begin{align}
        \mathcal{P}1:\  \max _{\{\mathbf{V}_{k}\}} \ & \sum_{k=1}^{K} \alpha_{k} R_{k} \\
        \text { s.t. }\ & \sum_{k=1}^{K}\left[\mathbf{V}_{k} \mathbf{V}_{k}^{H}\right]_{m, m} \leq P_{m}, \forall m \in \mathcal{M}, \\
        & R_{k} \geq r_{k}, \forall k,
    \end{align}
\end{subequations}
where $\alpha_{k}$ is a weighted scalar reflecting the priority of the user, $P_{m}$ denotes the transmit power budget of each transmit antenna, $r_{k}$ is the QoS constraint of each user. $[\mathbf{X}]_{m m}$ denotes the $m$-th diagonal element of $\mathbf{X}$, and $\mathcal{M}=\{1,2,...,N_{t}\}$ is the set of index of each transmit antenna. $R_{k}$ is the achievable rate of user $k$, which is given by
\begin{equation}
    \begin{aligned}
            R_{k}=\log \operatorname{det}
            & \bigg (\mathbf{I}+\mathbf{H}_{k} \mathbf{V}_{k} \mathbf{V}_{k}^{H} \mathbf{H}_{k}^{H}  \\
            & (\sum_{j \neq k}^{K} \mathbf{H}_{k} \mathbf{V}_{j} \mathbf{V}_{j}^{H} \mathbf{H}_{k}^{H}+\sigma^{2} \mathbf{I})^{-1} \bigg).
    \end{aligned}
\end{equation} 

It can be seen that $\mathcal{P}1$ is nonconvex, and it is hard to solve. Based on the well-known WMMSE framework, we transform the original problem into an equivalent problem and then solve it in the next section. 

\section{Proposed Solution}
In this section, we present the proposed solution to $\mathcal{P}1$. It contains two parts: i) reformulate the original problem based on the WMMSE framework, and ii) derive the solution to the reformulated problem by using ADMM algorithm.

\subsection{WMMSE-based Problem Transformation}
In this subsection, we first reformulate the original problem into an equivalent problem, which can be solved by BCD algorithm. Then we will present the formulation of each BCD steps. 
According to the WMMSE, we can obtain the following proposition.

\textit{Proposition 1:} The optimal solution to $\{\mathbf{V_{k}}\}$ of $\mathcal{P}1$ is equivalent to the problem
\begin{subequations}
    \begin{align}
        \mathcal{P}2:\  & \min _{\{\mathbf{W_{k}}, \mathbf{U_{k}}, \mathbf{V_{k}}\}} \ \sum_{k=1}^{K} \alpha_{k}\left(\operatorname{Tr}\left(\mathbf{W}_{k} \mathbf{E}_{k}\right)-\log \operatorname{det}\left(\mathbf{W}_{k}\right)-d\right) \\
        \text { s.t. }\ & \sum_{k=1}^{K}\left[\mathbf{V}_{k} \mathbf{V}_{k}^{H}\right]_{m, m} \leq P_{m}, \forall m, \\
        & \operatorname{Tr}\left(\mathbf{W}_{k} \mathbf{E}_{k}\right)-\log \operatorname{det}\left(\mathbf{W}_{k}\right)-d \leq-r_{k}, \forall k,
    \end{align}
\end{subequations}
where we recall that $d$ is the number of data streams per user, and $\mathbf{E}_{k}$ is the MSE matrix for user $k$, which is given by
\begin{equation}
    \begin{aligned}
        \mathbf{E}_{k}=
        & \left(\mathbf{I}-\mathbf{U}_{k}^{H} \mathbf{H}_{k} \mathbf{V}_{k}\right)\left(\mathbf{I}-\mathbf{U}_{k}^{H} \mathbf{H}_{k} \mathbf{V}_{k}\right)^{H} \\
        & +\sigma^{2} \mathbf{U}_{k}^{H} \mathbf{U}_{k}+\sum_{j \neq k}^{K} \mathbf{U}_{k}^{H} \mathbf{H}_{k} \mathbf{V}_{j} \mathbf{V}_{j}^{H} \mathbf{H}_{k}^{H} \mathbf{U}_{k}. \label{Ek}
    \end{aligned}
\end{equation}
$\mathbf{U_{k}}$ and $\mathbf{W_{k}}$ are auxiliary variables denoting the receiving matrix and the weight matrix of user $k$, respectively, and the optimal $\mathbf{U_{k}}$ and $\mathbf{W_{k}}$ for minimizing $\mathcal{P}2$ is given by
\begin{equation}
    \mathbf{U}_{k}=\left(\sum_{j=1}^{K} \mathbf{H}_{k} \mathbf{V}_{j} \mathbf{V}_{j}^{H} \mathbf{H}_{k}^{H}+\sigma_{k}^{2} \mathbf{I}\right)^{-1} \mathbf{H}_{k} \mathbf{V}_{k}, \forall k, \label{update_Uk}
\end{equation}
\begin{equation}
    \mathbf{W}_{k}=\left(\mathbf{I}-\mathbf{U}_{k}^{H} \mathbf{H}_{k} \mathbf{V}_{k}\right)^{-1}, \forall k
    \label{update_Wk},
\end{equation}
respectively.

\textit{Proof}: Please see Appendix A.

Since the objective function of $\mathcal{P}2$ is convex with respect to each of the optimization variables $\{\mathbf{W_{k}}, \mathbf{U_{k}}, \mathbf{V_{k}}\}$, we can apply the BCD algorithm to solve $\mathcal{P}2$ by repeating the following three steps until convergence.

i) Update $\{\mathbf{U}_{k}\}$ by equation \eqref{update_Uk},

ii) Update $\{\mathbf{W}_{k}\}$ by equation \eqref{update_Wk},

iii) Update $\{\mathbf{V}_{k}\}$ by solving following subproblem
\begin{equation}
    \begin{aligned}
        \mathcal{P}3:\ \min _{\{\mathbf{V}_{k}\}} \  & \alpha_{k} \sum_{k=1}^{K} \operatorname{Tr}\left(\mathbf{W}_{k} \mathbf{E}_{k}\right) \\
        \text { s.t. }\ & \sum_{k=1}^{K}\left[\mathbf{V}_{k} \mathbf{V}_{k}^{H}\right]_{m, m} \leq P_{m}, \forall m, \\
        & \operatorname{Tr}\left(\mathbf{W}_{k} \mathbf{E}_{k}\right) \leq \log \operatorname{det}\left(\mathbf{W}_{k}\right)+d-r_{k}, \forall k.
    \end{aligned}
\end{equation}

The subproblem $\mathcal{P}3$ is hard to solve since variables $\{\mathbf{V}_{k}\}$ are coupled in both QoS constraints and PAPCs. In the next subsection, we will address the subproblem using the ADMM algorithm.

\subsection{The Proposed ADMM for Solving $\mathcal{P}3$}
In this subsection, we will first introduce auxiliary variables to decouple the QoS constraints and PAPCs, so that we can solve the problem with ADMM algorithm. Next, we will provide the closed-form solution for each subproblem at each ADMM iteration. 

\subsubsection{Problem Decomposition}
Replacing the optimal $\{\mathbf{U_{k}}\}, \{\mathbf{W_{k}}\}$ in \eqref{update_Uk} and \eqref{update_Wk} into $\mathcal{P}3$ and introducing auxiliary variables $\mathbf{X}_{k, j}=\mathbf{U}_{k}^{H} \mathbf{H}_{k} \mathbf{V}_{j}, \forall k, \forall j$, we obtain the following problem with respect to $\{\mathbf{V}_{k}\}$ and $\{\mathbf{X}_{k,j}\}$.

\begin{equation}
    \begin{aligned}
    \ \min _{\{\mathbf{V}_{k}, \mathbf{X}_{k, j}\}} & \sum_{k=1}^{K} \alpha_{k} \operatorname{Tr}\left(-\mathbf{W}_{k} \mathbf{X}_{k, k}^{H}-\mathbf{W}_{k} \mathbf{X}_{k, k}\right)+ \\
    & \sum_{k=1}^{K} \operatorname{Tr}\left(\sum_{j=1}^{K} \alpha_{j} \mathbf{W}_{j} \mathbf{X}_{j, k} \mathbf{X}_{j, k}^{H}\right) \\
    \text { s.t.} \ & \operatorname{Tr}\left(\mathbf{W}_{k} \mathbf{X}_{k, k} \mathbf{X}_{k, k}^{H}-\mathbf{W}_{k} \mathbf{X}_{k, k}^{H}-\mathbf{W}_{k} \mathbf{X}_{k, k}\right)+ \\
    & \operatorname{Tr}\left(\mathbf{W}_{k}+\sigma^{2} \mathbf{W}_{k} \mathbf{U}_{k}^{H} \mathbf{U}_{k}+\mathbf{W}_{k}\left(\sum_{j \neq k} \mathbf{X}_{k, j} \mathbf{X}_{k, j}^{H}\right)\right) \\
    & \leq \log \operatorname{det}\left(\mathbf{W}_{k}\right)+d-r_{k}, \forall k, \\
    & \sum_{k=1}^{K}\left[\mathbf{V}_{k} \mathbf{V}_{k}^{H}\right]_{m, m} \leq P_{m}, \forall m, \\
    & \mathbf{X}_{k, j}=\mathbf{U}_{k}^{H} \mathbf{H}_{k} \mathbf{V}_{j}, \forall k, \forall j. \\
    \label{p4}
    \end{aligned} 
\end{equation}

By placing the equality constraints $\mathbf{X}_{k, j}=\mathbf{U}_{k}^{H} \mathbf{H}_{k} \mathbf{V}_{j}$ into the augmented Lagrangian function of \eqref{p4},  we obtain
\begin{equation}
    \begin{aligned}
        \mathcal{L} \ & \left(\left\{\mathbf{V}_{k}\right\},\left\{\mathbf{X}_{k, j}\right\}\right) \\
        = \ & \sum_{k=1}^{K} \alpha_{k} \operatorname{Tr}\left(-\mathbf{W}_{k} \mathbf{X}_{k, k}^{H}-\mathbf{W}_{k} \mathbf{X}_{k, k}\right) \\
        + \ & \sum_{k=1}^{K} \operatorname{Tr}\left(\sum_{j=1}^{K} \alpha_{j} \mathbf{W}_{j} \mathbf{X}_{j, k} \mathbf{X}_{j, k}^{H}\right)  \\
        + \ & \frac{\rho}{2} \sum_{k=1}^{K} \sum_{j=1}^{K}\left\|\mathbf{U}_{k}^{H} \mathbf{H}_{k} \mathbf{V}_{j}-\mathbf{X}_{k, j}+\mathbf{\lambda}_{k, j}\right\|^{2}, \\
    \end{aligned}
\end{equation}
where $\mathbf{\lambda}_{k, j}$ are dual variables associated with the equality constraints $\mathbf{X}_{k, j}=\mathbf{U}_{k}^{H} \mathbf{H}_{k} \mathbf{V}_{j}, \forall k, \forall j$, and $\rho$ is the penalty parameter. 

According to the convergence requirements of the ADMM algorithm, we split the primal variables into two blocks $\{\mathbf{V}_{k}\}$ and $\{\mathbf{X}_{k, j}\}$. Therefore, at the $t$-th iteration, the algorithm iteratively solves the following three problems:

\paragraph{} Optimizing $\mathbf{V}_{k}(t)$ with fixed $\mathbf{X}_{k, j}(t-1)$ and $\mathbf{\lambda}_{k, j}(t-1)$:
\begin{equation}
    \begin{aligned}
        \min _{\left\{\mathbf{V}_{k}\right\}} \ & \mathcal{L}\left(\left\{\mathbf{V}_{k}\right\},\left\{\mathbf{X}_{k, j}\right\}\right) \\
        \text { s.t. } & \sum_{k=1}^{K}\left[\mathbf{V}_{k} \mathbf{V}_{k}^{H}\right]_{m, m} \leq P_{m}, \forall m. \label{update_Vk}
    \end{aligned}
\end{equation}

\paragraph{} Optimizing $\mathbf{X}_{k, j}(t)$ with fixed $\mathbf{V}_{k}(t)$ and $\mathbf{\lambda}_{k, j}(t-1)$:
\begin{equation}
    \begin{aligned}
        \min _{\left\{\mathbf{X}_{k, j}\right\}} \ & \mathcal{L}\left(\left\{\mathbf{V}_{k}\right\},\left\{\mathbf{X}_{k, j}\right\}\right) \\
        \text { s.t. } \ & \operatorname{Tr}\left(\mathbf{W}_{k} \mathbf{X}_{k, k} \mathbf{X}_{k, k}^{H}-\mathbf{W}_{k} \mathbf{X}_{k, k}^{H}-\mathbf{W}_{k} \mathbf{X}_{k, k}\right) \\
        + \ & \operatorname{Tr}\left(\mathbf{W}_{k}\left(\sum_{j \neq k} \mathbf{X}_{k, j} \mathbf{X}_{k, j}^{H}\right)\right) \leq e_{k}, \forall k. \label{update_X}
    \end{aligned}
\end{equation}
where
\begin{equation}
    e_{k}=\log \operatorname{det}\left(\mathbf{W}_{k}\right)+d-r_{k}-\operatorname{Tr}\left(\mathbf{W}_{k}+\sigma^{2} \mathbf{W}_{k} \mathbf{U}_{k}^{H} \mathbf{U}_{k}\right).
\end{equation}

\paragraph{} Update $\mathbf{\lambda}_{k, j}(t)$ with fixed $\mathbf{V}_{k}(t)$ and $\mathbf{X}_{k, j}(t)$ by
\begin{equation}
    \mathbf{\lambda}_{k, j}(t)=\mathbf{\lambda}_{k, j}(t-1)+\rho\left(\mathbf{U}_{k}^{H} \mathbf{H}_{k} \mathbf{V}_{j}-\mathbf{X}_{k, j}\right), \forall k, \forall j. \label{update_lambda}
\end{equation}
It should be noted that for notational simplicity, the iteration index $t$ is omitted in the following.

\subsubsection{Solution to Problem \eqref{update_Vk}}
The problem \eqref{update_Vk} is given by
\begin{equation}
    \begin{aligned}
        \min _{\left\{\mathbf{V}_{k}\right\}} \ & \frac{\rho}{2} \sum_{k=1}^{K} \sum_{j =1}^{K}\left\|\mathbf{U}_{k}^{H} \mathbf{H}_{k} \mathbf{V}_{j}-\mathbf{X}_{k, j}+\mathbf{\lambda}_{k, j}\right\|^{2} \\
        \text { s.t. } \  & \sum_{k=1}^{K}\left[\mathbf{V}_{k} \mathbf{V}_{k}^{H}\right]_{m, m} \leq P_{m}, \forall m. \label{first_subproblem}
    \end{aligned}
\end{equation}

To derive the closed-form updates of the problem, we reformulate the PAPCs to make it more tractable. We denote $\mathbf{h}_{k, m} \in \mathbb{C}^{N_{r} \times 1}$ as the $m$-th column of $\mathbf{H}_{k}$ and $\mathbf{v}_{k, m} \in \mathbb{C}^{d \times 1}$ as the $m$-th column of $\mathbf{V}_{k}^{H}$. Thus, it is obvious that
\begin{equation}
    \mathbf{H}_{j} \mathbf{V}_{k}=\sum_{m=1}^{N_{t}} \mathbf{h}_{j, m} \mathbf{v}_{k, m}^{H}.
\end{equation}
Moreover, the PAPCs can be reformulated as 
\begin{equation}
    \sum_{k=1}^{K}\left[\mathbf{V}_{k} \mathbf{V}_{k}^{H}\right]_{m, m}=\sum_{k=1}^{K} \operatorname{Tr}\left(\mathbf{v}_{k, m} \mathbf{v}_{k, m}^{H}\right), \forall m.
\end{equation}
Therefore, the subproblem \eqref{first_subproblem} can be reformulated as
\begin{equation}
    \begin{aligned}
        \min _{\left\{\mathbf{v}_{k,m}\right\}} \ & \frac{\rho}{2} \sum_{k=1}^{K} \sum_{j=1}^{K}\left\|\mathbf{U}_{j}^{H} \sum_{m=1}^{N_{t}} \mathbf{h}_{j, m} \mathbf{v}_{k, m}^{H}-\mathbf{X}_{j, k}+\mathbf{\lambda}_{j, k}\right\|^{2} \\
        \text { s.t. } \ & \sum_{k=1}^{K} \operatorname{Tr}\left(\mathbf{v}_{k, m} \mathbf{v}_{k, m}^{H}\right) \leq P_{m}, \forall m.
    \end{aligned}
\end{equation}
By using the Lagrangian multiplier method, the optimal $\mathbf{v}_{k,m}$ is given by
\begin{equation}
    \begin{aligned}
        & \mathbf{v}_{k, m}^{opt}(\mu_{m})=\left(\rho \sum_{j=1}^{K} \mathbf{h}_{j, m}^{H} \mathbf{U}_{j} \mathbf{U}_{j}^{H} \mathbf{h}_{j, m}+2 \mu_{m}\right)^{-1} \\
        & \left(\rho \sum_{j=1}^{K} \mathbf{X}_{j, k}^{H} \mathbf{U}_{j}^{H} \mathbf{h}_{j, m}-\rho \sum_{j=1}^{K} \lambda_{j, k}^{H} \mathbf{U}_{j}^{H} \mathbf{h}_{j, m} \right. \\
        & \left. -\rho \sum_{j=1}^{K}\left(\sum_{n \neq m}^{N_{t}} \mathbf{v}_{k, n} \mathbf{h}_{j, n}^{H}\right) \mathbf{U}_{j} \mathbf{U}_{j}^{H} \mathbf{h}_{j, m}\right), \forall k, \forall m,
        \label{vkm}
    \end{aligned}
\end{equation}
where $\mu_{m} \geq 0, \forall m$ is the Lagrange multiplier. Thus, its solution has two cases. If $\sum_{k=1}^{K} \operatorname{Tr}\left(\mathbf{v}_{k, m}(0) \mathbf{v}_{k, m}^{H}(0)\right) \leq P_{m}$, then $\mathbf{v}_{k, m}^{opt}(\mu_{m})=\mathbf{v}_{k, m}(0)$. Otherwise, we must have 
\begin{equation}
    \sum_{k=1}^{K} \operatorname{Tr}\left(\mathbf{v}_{k, m}\left(\mu_{m}\right) \mathbf{v}_{k, m}^{H}\left(\mu_{m}\right)\right)=P_{m}, \forall m. \label{um}
\end{equation}
Substituting \eqref{vkm} into \eqref{um}, we can obtain a decreasing function with respect to $\mu_{m}$, which is given by
\begin{equation}
    \frac{1}{\left(q+2 \mu_{m}\right)^{2}} \operatorname{Tr}(\mathbf{D})=P_{m},
\end{equation}
where $q = \rho \sum_{j=1}^{K} \mathbf{h}_{j, m}^{H} \mathbf{U}_{j} \mathbf{U}_{j}^{H} \mathbf{h}_{j, m}$, and $\mathbf{D}=\sum_{k=1}^{K} \mathbf{D}_{k} \mathbf{D}_{k}^{H}$, 
\begin{equation}
    \begin{aligned}
        \mathbf{D}_{k}=
        & \rho \sum_{j=1}^{K} \mathbf{X}_{j, k}^{H} \mathbf{U}_{j}^{H} \mathbf{h}_{j, m}-\rho \sum_{j=1}^{K} \lambda_{j, k}^{H} \mathbf{U}_{j}^{H} \mathbf{h}_{j, m} \\
        & -\rho \sum_{j=1}^{K}\left(\sum_{n \neq m}^{N_{\mathrm{t}}} \mathbf{v}_{k, n} \mathbf{h}_{j, n}^{H}\right) \mathbf{U}_{j} \mathbf{U}_{j}^{H} \mathbf{h}_{j, m}.
    \end{aligned}
\end{equation}
Therefore, we can find the $\mu_{m}^{*}$ using bisection method. Then the optimal $\mathbf{v}_{k, m}^{opt}$ can be obtained according to \eqref{vkm}. We recall that $\mathbf{v}_{k, m} \in \mathbb{C}^{d \times 1}$ is the $m$-th column of $\mathbf{V}_{k}^{H}$, thus we can map $\mathbf{v}_{k, m}^{opt}$ to $\mathbf{V}_{k}^{opt}$ using the relationship between them.

    
\subsubsection{Solution to Problem \eqref{update_X}}
The problem \eqref{update_X} can be rewritten as
\begin{equation}
    \begin{aligned}
        \min _{\left\{\mathbf{X}_{k, j}\right\}} \ & \sum_{k=1}^{K} \alpha_{k} \operatorname{Tr}\left(-\mathbf{W}_{k} \mathbf{X}_{k, k}^{H}-\mathbf{W}_{k} \mathbf{X}_{k, k}\right) \\
        + \ & \sum_{k=1}^{K} \operatorname{Tr}\left(\sum_{j=1}^{K} \alpha_{j} \mathbf{W}_{j} \mathbf{X}_{j, k} \mathbf{X}_{j, k}^{H}\right) \\
        + \ & \frac{\rho}{2} \sum_{k=1}^{K} \sum_{j=1}^{K}\left\|\mathbf{U}_{k}^{H} \mathbf{H}_{k} \mathbf{V}_{j}-\mathbf{X}_{k, j}+\lambda_{k, j}\right\|^{2} \\
        \text { s.t. } \ & \operatorname{Tr}\left(\mathbf{W}_{k} \mathbf{X}_{k, k} \mathbf{X}_{k, k}^{H}-\mathbf{W}_{k} \mathbf{X}_{k, k}^{H}-\mathbf{W}_{k} \mathbf{X}_{k, k}\right) \\
        + \ & \operatorname{Tr}\left(\mathbf{W}_{k}\left(\sum_{j \neq k}^{K} \mathbf{X}_{k, j} \mathbf{X}_{k, j}^{H}\right)\right) \leq e_{k}, \forall k. \\ \label{second subproblem}
    \end{aligned}
\end{equation}

The problem \eqref{second subproblem} can be decomposed into $K$ independent subproblem, and the Lagrange function for $k$-th subproblem is given by
\begin{equation}
    \begin{aligned}
        & \mathcal{L}\left(\left\{\mathbf{X}_{k, j}\right\}_{j=1}^{K}, \tau_{k}\right)=\alpha_{k} \operatorname{Tr}\left(-\mathbf{W}_{k} \mathbf{X}_{k, k}^{H}-\mathbf{W}_{k} \mathbf{X}_{k, k}\right) \\
        & +\sum_{j \neq k}^{K} \operatorname{Tr}\left(\alpha_{k} \mathbf{W}_{k} \mathbf{X}_{k, j} \mathbf{X}_{k, j}^{H}\right)+\operatorname{Tr}\left(\alpha_{k} \mathbf{W}_{k} \mathbf{X}_{k, k} \mathbf{X}_{k, k}^{H}\right) \\
        & +\frac{\rho}{2} \sum_{j=1}^{K}\left\|\mathbf{U}_{k}^{H} \mathbf{H}_{k} \mathbf{V}_{j}-\mathbf{X}_{k, j}+\mathbf{\lambda}_{k, j}\right\|^{2} \\
        & +\tau_{k} 
        \bigg (\operatorname{Tr}\left(\mathbf{W}_{k} \mathbf{X}_{k, k} \mathbf{X}_{k, k}^{H}-\mathbf{W}_{k} \mathbf{X}_{k, k}^{H}-\mathbf{W}_{k} \mathbf{X}_{k, k}\right) \\
        & +\operatorname{Tr}\big(\mathbf{W}_{k}\big(\sum_{j \neq k}^{K} \mathbf{X}_{k, j} \mathbf{X}_{k, j}^{H}\big)\big)-e_{k} \bigg).
    \end{aligned}
\end{equation}
Thus, the first-order optimality condition with respect to $\mathbf{X}_{k, j}$ is given by
\begin{equation}
    \begin{aligned}
        \mathbf{X}_{k, j}^{o p t}(\tau_{k})
        =&\left(2 \alpha_{k} \mathbf{W}_{k}+\rho \mathbf{I}+2 \tau_{k} \mathbf{W}_{k}\right)^{-1} \\
        &\left(\rho \mathbf{U}_{k}^{H} \mathbf{H}_{k} \mathbf{V}_{j}+\rho \mathbf{\lambda}_{k, j}\right), 
        \label{Xkj}
    \end{aligned}
\end{equation}
where $\ k \neq j$, and

\begin{equation}
    \begin{aligned}
        \mathbf{X}_{k, k}^{opt}(\tau_{k})= 
        & \left(2 \alpha_{k} \mathbf{W}_{k}+\rho \mathbf{I}+2 \tau_{k} \mathbf{W}_{k}\right)^{-1} \\
        & \left(2 \alpha_{k} \mathbf{W}_{k}+\rho \mathbf{\lambda}_{k, k}+\rho \mathbf{U}_{k}^{H} \mathbf{H}_{k} \mathbf{V}_{k}+2 \tau_{k} \mathbf{W}_{k}\right), \label{Xkk}
    \end{aligned}
\end{equation}
where $\tau_{k}$ is the Lagrange multiplier that satisfies $\tau_{k} \geq 0$, and we give the solution to obtain the optimal $\tau_{k}^{*}$. 

\textit{Lemma 1}: the optimal $\tau_{k}^{*}$ can be obtained by bisection search.

\textit{Proof}: Please see Appendix B.

\subsection{Overall Algorithm and Complexity Analysis}
The overall algorithm is presented in $\mathbf{Algorithm}$ \ref{alg:algorithm1}, where $L_{o}$ is the number of the outer BCD iterations and $L_{i}$ is the number of inner ADMM iterations.

\begin{algorithm}[H]
    \caption{The Proposed Solution to $\mathcal{P}1$}
    \label{alg:algorithm1}
    \begin{algorithmic}[1] 
        \STATE {Initialize $\{\mathbf{V}_{k}\}$ such that $\sum_{k=1}^{K}\left[\mathbf{V}_{k} \mathbf{V}_{k}^{H}\right]_{m, m} \leq P_{m}, \forall m$. Set the outer iteration index $t_o=0$}.
        \REPEAT
        \STATE {1). Update $\{\mathbf{U}_{k}\}$ by equation \eqref{update_Uk}}.
        \STATE {2). Update $\{\mathbf{W}_{k}\}$ by equation \eqref{update_Wk}}.
        \REPEAT
        \STATE {Set the inner iteration index $t_i=0$}.
        \STATE {1). Update $\{\mathbf{V}_{k}\}$ by solving \eqref{update_Vk}}.
        \STATE {2). Update $\{\mathbf{X}_{k, j}\}$ by solving \eqref{update_X}}.
        \STATE {3). Update $\{\mathbf{\lambda}_{k, j}\}$ by solving \eqref{update_lambda}}.
        \STATE {$t_i = t_i + 1$}.
        \UNTIL{The objective function converges or $t_i \geq L_{i}$.}
        \STATE {$t_o = t_o + 1$}.
        \UNTIL{The objective function converges or $t_o \geq L_{o}$.}
    \end{algorithmic} 
\end{algorithm} 

Now we analyze the complexity of the proposed algorithm. We focus on the multiplication of the matrices and ignore the bisection search steps since its complexity is much more less. The complexity of updating $\{\mathbf{U}_{k}\}$ and  $\{\mathbf{W}_{k}\}$ per iteration is $\mathcal{O}\left(K^{2} N_{r}^{2} N_{t}\right)$ and $\mathcal{O}\left(K N_{r}^{3}\right)$ according to \eqref{update_Uk} and \eqref{update_Wk}, respectively. In the inner ADMM iteration, the complexity of updating $\{\mathbf{V}_{k}\}$ per iteration is $\mathcal{O}\left(K^{2} N_{r}^{2} N_{t}^{2}\right)$. Meanwhile, the complexity of updating $\mathbf{X}_{k, j}$ and $\mathbf{\lambda}_{k, j}$ per iteration are both $\mathcal{O}\left(K^{2} N_{r}^{2} N_{t}\right)$. Thus, the complexity of the algorithm is
$\mathcal{O}\left(L_{o}\left(K^{2} N_{r}^{2} N_{t}+K N_{r}^{3}+L_{i}\left(K^{2} N_{r}^{2} N_{t}^{2} + 2K^{2} N_{r}^{2} N_{t}\right)\right)\right)$
, which is polynomial.

\section{Simulation Results}
In this section, we present the simulation results to validate the effectiveness of the proposed algorithm. The simulation settings are as follows unless otherwise stated. The BS is equipped with $N_{t}=16$ transmit antennas, and there are 4 users each equipped with $N_{r}=2$ receive antennas. The number of data streams per user are set to $d=2$. The power budget of each BS antenna is $P_{m}=P_{max}/N_{t}$ where $P_{max}=10$ dBm. The channel matrix $\mathbf{H}_{k}$ from the BS to the user $k$ is generated by $\mathbf{H}_{k}=\sqrt{\theta_{k}} \mathbf{H}_{k}^{s}$. $\theta_{k}$ is the pathloss given by $128.1+37.6lg (d_k)$ dB, where $d_k$ is the distance between the BS and the user $k$ taking range from [0.1, 0.2] km. The small-scale Rayleigh fading $\mathbf{H}_{k}^{s}$ follows complex Gaussian distribution $\mathcal{C} \mathcal{N}(0,1)$. The priority weights $\alpha_{k}, \forall k$ of the users are set to be equal. The penalty parameter $\rho$ is 1.

\begin{figure}[htbp]
    \centerline{\includegraphics[width=6cm ,trim=5 0 20 10,clip]{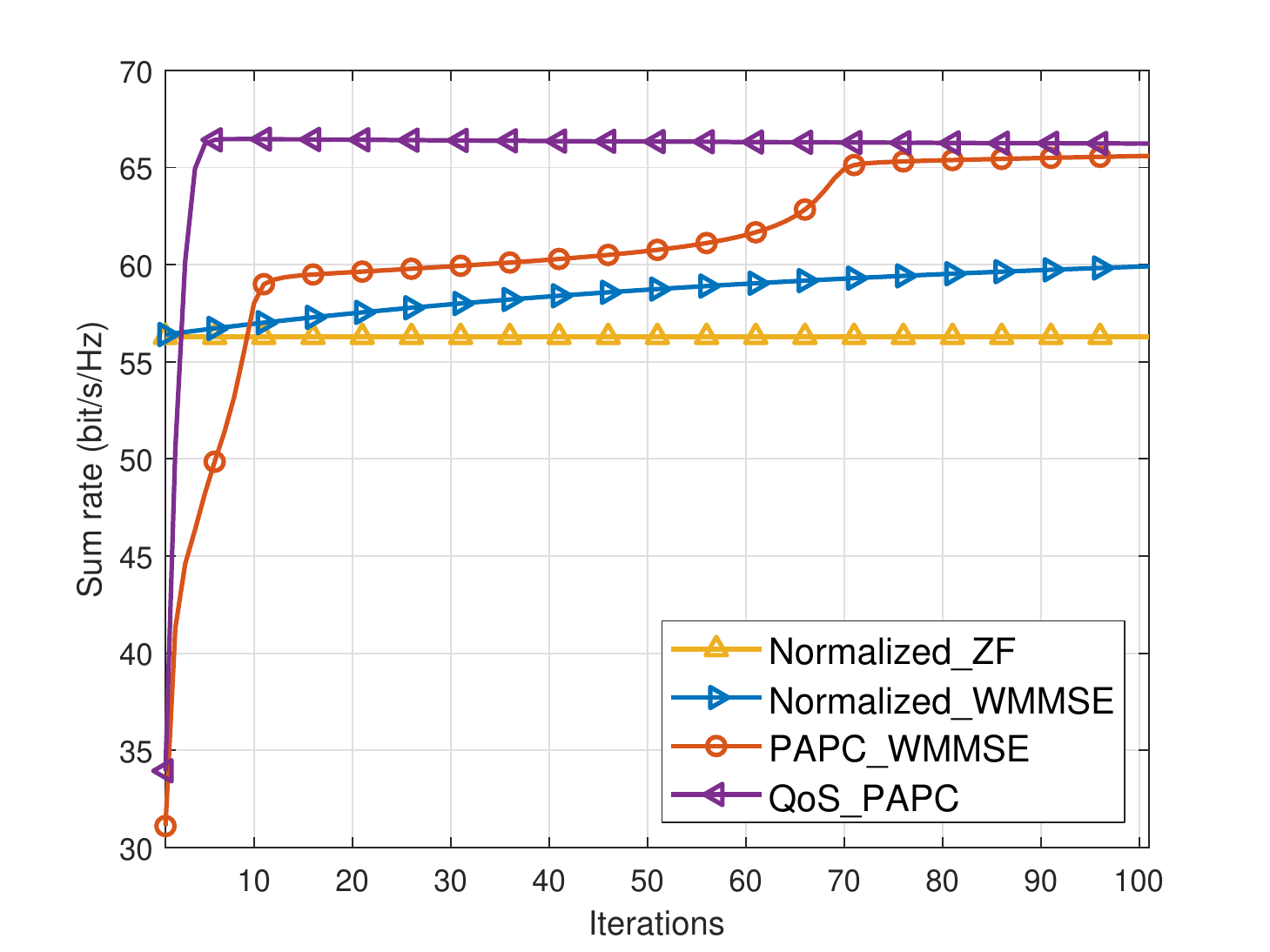}}
    \caption{Convergence of the proposed QoS\_PAPC algorithm.}
    \label{fig_iters}
\end{figure}

First, we validate the convergence performance of the proposed algorithm in Fig.~\ref{fig_iters}. For comparison, we consider three benchmarks, i.e., Normalize\_ZF, Normalized\_WMMSE and PAPC\_WMMSE. The Normalize\_ZF and the Normalized\_WMMSE are given by normalizing the solution of ZF and WMMSE to satisfy the PAPCs, respectively.
The PAPC\_WMMSE is designed to directly solve the WSR problem with PAPCs \cite{rethinking}. Here, the QoS rate constraint of each user is set to 0 to exam the convergence of the proposed method. It can be seen in Fig.~\ref{fig_iters} that the proposed method converges to the same value as the PAPC\_WMMSE. It also can be seen that the proposed method takes fewer outer BCD iterations than PAPC\_WMMSE. This validates the convergence and efficiency of the proposed algorithm. The Normalized\_WMMSE shows degraded performance since directly scaling the solution with SPC constraint to satisfy PAPCs can not fully exploit the power budget of each transmit antenna. Similarly, the Normalized\_ZF yields worst performance since it is a heuristic algorithm and it does not consider the objective of WSR maximization. 

\begin{figure}[htbp]
    \centerline{\includegraphics[width=6cm ,trim=5 0 20 10,clip]{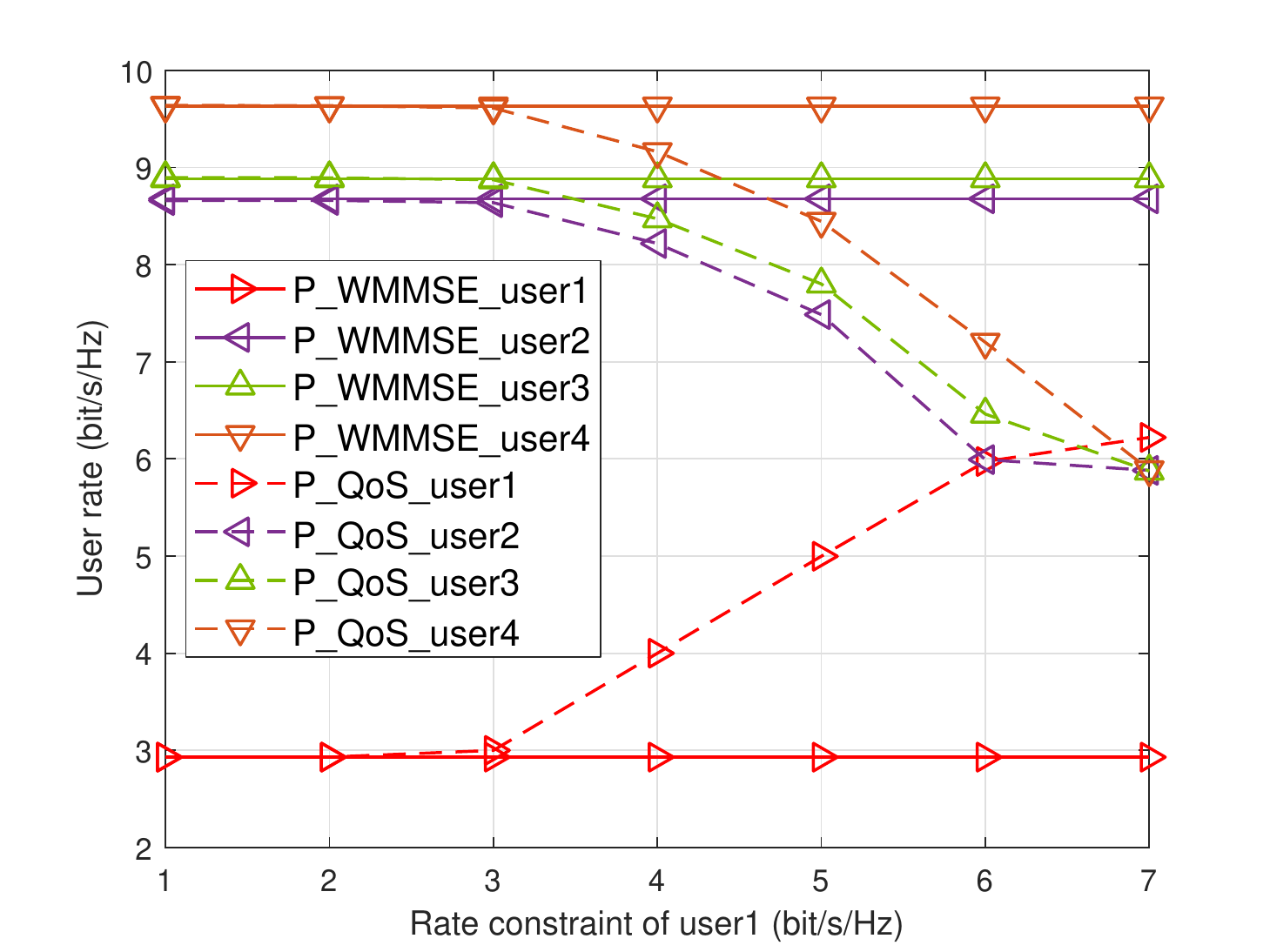}}
    \caption{Rate of each user vs. QoS constraint change of user 1: $N_{t}=16$.}
    \label{fig_user_rate_16}
\end{figure}

To evaluate the effectiveness of the proposed algorithm, we consider a scenario in which the position of 4 users are fixed. We fix the QoS rate constraints of the last three users with 6 bit/s/Hz and change the QoS constraint of the first user from 1 bit/s/Hz to 7 bit/s/Hz. We present the rate of each user in Fig.~\ref{fig_user_rate_16}. \textit{P\_WMMSE\_useri} denotes the rate of user $i$ computed by PAPC\_WMMSE and \textit{P\_qos\_useri} denotes the rate of user $i$ computed by the proposed QoS\_PAPC, $i=1,2,3,4$. We can observe that the rates of remaining three users computed by PAPC\_QoS decrease while the rate of user 1 increases so that all users can satisfy the QoS constraints, while the rates by PAPC\_WMMSE remain the same. However,  we note that when the rate constraint of user 1 becomes higher, i.e., 7 bit/s/Hz, the rate of user 1 computed by PAPC\_QoS can not meet the QoS requirements either. This is because the problem becomes infeasible with given limited power budget.

\begin{figure}[htbp]
    \centerline{\includegraphics[width=6cm ,trim=5 0 20 10,clip]{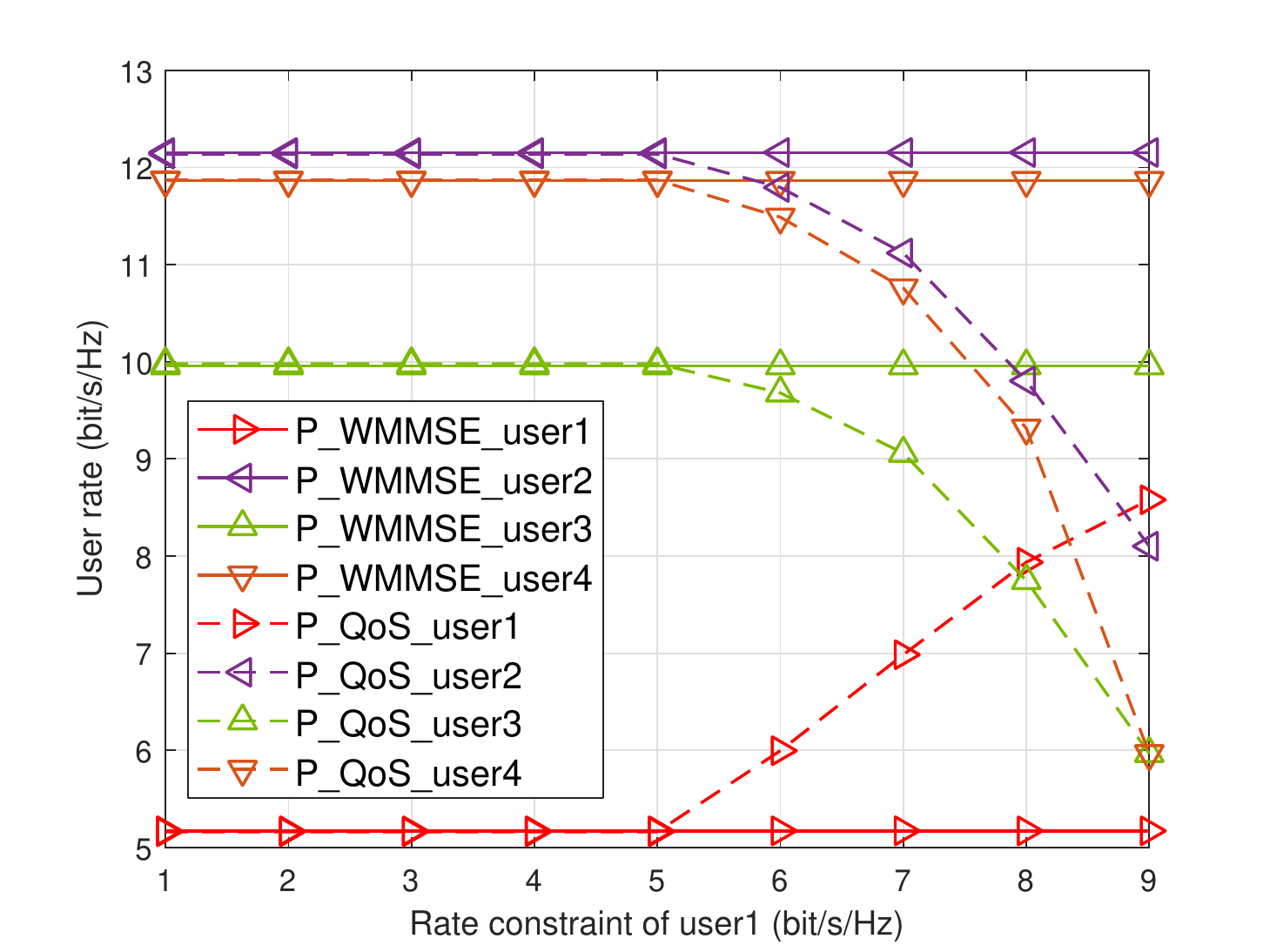}}
    \caption{Rate of each user vs. QoS constraint change of user 1:$N_{t}=32$.}
    \label{fig_user_rate_32}
    \vspace{-1.5em}
\end{figure}

Fig.~\ref{fig_user_rate_32} shows the rate of each user under different QoS constraints with 32 transmit antennas. It can be seen that when there are more transmit antennas, the rate of each user becomes higher. Similarly, the proposed method can adjust precoding matrices according to the QoS constraints of users to allow all users satisfy the QoS constraints, while the rates computed by PAPC\_WMMSE remain the same since it does not take QoS constraints into consideration.

\section{Conclusion}
In this paper, we proposed a downlink precoding algorithm that maximizes the weighted sum rate while considering both QoS constraints and PAPCs. We transformed the original problem into an equivalent problem which can be solved by using the BCD algorithm. The algorithm iteratively updates three subproblems. To deal with the coupled variables among constraints, we adopt the ADMM algorithm to decouple the QoS constraints and PAPCs. Simulation results demonstrate that the proposed method converges faster than benchmarks and is able to adjust precoding matrices to make users satisfy the QoS constraints with power constraints.

\begin{appendices}
\vspace{-0.5em}
    \section{Proof of Proposition 1}
    \vspace{-0.5em}
    With fixed $\{\mathbf{V_{k}}, \mathbf{W_{k}}\}$, the first-order optimality condition of $\mathcal{P}2$ with respect to $\{\mathbf{U_{k}}\}$ is given by \eqref{update_Uk}. Substituting  \eqref{update_Uk} into \eqref{Ek}, we obtain $\mathbf{E}_{k}=\mathbf{I}-\mathbf{U}_{k}^{H} \mathbf{H}_{k} \mathbf{V}_{k}, \forall k$. Besides, by checking the first order optimality condition for $\{\mathbf{W_{k}}\}$, the optimal $\{\mathbf{W_{k}}\}$ in $\mathcal{P}2$ with variables $\{\mathbf{V_{k}}, \mathbf{U_{k}}\}$ fixed is given by $\mathbf{W}_{k}^{\text {opt }}=\mathbf{E}_{k}^{-1}$. Substituting the optimal $\{\mathbf{U_{k}}, \mathbf{W_{k}}\}$ into $\mathcal{P}2$, we can obtain the equivalent problem:
    \begin{equation}
        \begin{aligned}
            \max _{\left\{\mathbf{V}_{k}\right\}} & \sum_{k=1}^{K} \alpha_{k} \log \operatorname{det}\left(\mathbf{E}_{k}^{-1}\right) \\
            \text { s.t. } & \sum_{k=1}^{K}\left[\mathbf{V}_{k} \mathbf{V}_{k}^{H}\right]_{m, m} \leq P_{m}, \forall m, \\
            & \log \operatorname{det}\left(\mathbf{E}_{k}^{-1}\right) \geq r_{k}, \forall k. \label{appendix_A}
        \end{aligned}
    \end{equation}
    Defining $\mathbf{J}_{k}=\sum_{j}^{K} \mathbf{H}_{k} \mathbf{V}_{j} \mathbf{V}_{j}^{H} \mathbf{H}_{k}^{H}+\sigma^{2} \mathbf{I}$, we can obtain $\mathbf{E}_{k}=\mathbf{I}-\mathbf{V}_{k}^{H} \mathbf{H}_{k}^{H} \mathbf{J}_{k} \mathbf{H}_{k} \mathbf{V}_{k}$. And
    \begin{equation}
        \begin{aligned}
            \log \operatorname{det}\left(\mathbf{E}_{k}^{-1}\right)
            & =\log \operatorname{det}\left(\left(\mathbf{I}-\mathbf{V}_{k}^{H} \mathbf{H}_{k}^{H} \mathbf{J}_{k} \mathbf{H}_{k} \mathbf{V}_{k}\right)^{-1}\right) \\
            & =\log \operatorname{det}\left(\mathbf{I}+\mathbf{V}_{k}^{H} \mathbf{H}_{k}^{H} \mathbf{C}_{k}^{-1} \mathbf{H}_{k} \mathbf{V}_{k}\right) \\
            & =\log \operatorname{det}\left(\mathbf{I}+\mathbf{H}_{k} \mathbf{V}_{k} \mathbf{V}_{k}^{H} \mathbf{H}_{k}^{H} \mathbf{C}_{k}^{-1}\right), \label{logdetE}
        \end{aligned}
    \end{equation} 
    where $\mathbf{C}_{k}=\sum_{j \neq k}^{K} \mathbf{H}_{k} \mathbf{V}_{j} \mathbf{V}_{j}^{H} \mathbf{H}_{k}^{H}+\sigma^{2} \mathbf{I}$. Particularly, the second equality is obtained by Woodbury matrix identity and the third equality is given by $\operatorname{det}\left(\mathbf{I}+\mathbf{A}_{1} \mathbf{A}_{2}\right)=\operatorname{det}\left(\mathbf{I}+\mathbf{A}_{2} \mathbf{A}_{1}\right)$. Substituting \eqref{logdetE} into \eqref{appendix_A}, we find that the problem \eqref{appendix_A} is equal to $\mathcal{P}1$, thus the proof is completed.

    \vspace{-0.5em}
    \section {Proof of Lemma 1}
    \vspace{-0.5em}
    Same as $\mu_{m}$, the optimal value of $\tau_{k}$ has two cases. If 
    \begin{equation}
        \begin{aligned}
            &\operatorname{Tr}\left(-\mathbf{W}_{k} \mathbf{X}_{k, k}^{H}(0)-\mathbf{W}_{k} \mathbf{X}_{k, k}(0)\right) \\
            &+ \operatorname{Tr}\left(\mathbf{W}_{k}\left(\sum_{j=1}^{K} \mathbf{X}_{k, j}(0) \mathbf{X}_{k, j}^{H}(0)\right)\right) \leq e_{k},
        \end{aligned}
    \end{equation}
    then $\mathbf{X}_{k, j}^{opt}(\tau_{k})=\mathbf{X}_{k, j}(0)$. Otherwise, we must have 
    \begin{equation}
        \begin{aligned}
            &\operatorname{Tr}\left(-\mathbf{W}_{k} \mathbf{X}_{k, k}^{H}(\tau_{k})-\mathbf{W}_{k} \mathbf{X}_{k, k}(\tau_{k})\right) \\
            &+ \operatorname{Tr}\left(\mathbf{W}_{k}\left(\sum_{j=1}^{K} \mathbf{X}_{k, j}(\tau_{k}) \mathbf{X}_{k, j}^{H}(\tau_{k})\right)\right) = e_{k}. \label{equality}
        \end{aligned}
    \end{equation}
    Since matrix $\mathbf{W}_{k}$ is a positive semidefinit hermitian matrix \cite{wmmse}, we perform the eigendecomposition and obtain $\mathbf{W}_{k}=\mathbf{D}_{k} \mathbf{\Lambda}_{k} \mathbf{D}_{k}^{H}$, where $\mathbf{D}_{k}$ is an unitary matrix satisfying $\mathbf{D}_{k} \mathbf{D}_{k}^{H}=\mathbf{I}$ and $\mathbf{\Lambda}_{k}$ is a diagonal matrix with non-negative diagonal elements. Substituting \eqref{Xkj} and \eqref{Xkk} into \eqref{equality}, we obtain
    \begin{equation}
        \begin{aligned}
            & \sum_{m=1}^{d} \frac{[\mathbf{\Lambda}_{k}]_{m m}\left[\mathbf{D}_{k}^{H} \mathbf{\Phi}_{k} \mathbf{D}_{k}\right]_{m m}}{\left(2 \alpha_{k}[\mathbf{\Lambda}_{k}]_{m m}+\rho+2 \tau_{k}[\mathbf{\Lambda}_{k}]_{m m}\right)^{2}} \\
            & +\sum_{m=1}^{d}\left(\frac{[\mathbf{\Lambda_{k}}]_{m m}\left[\mathbf{G}_{k}\right]_{m m}}{\left(2 \alpha_{k}[\mathbf{\Lambda}_{k}]_{m m}+\rho+2 \tau_{k}[\mathbf{\Lambda}_{k}]_{m m}\right)^{2}}-[\mathbf{\Lambda}_{k}]_{m m}\right)=e_{k}. \\ \label{equation_tau}
        \end{aligned}
    \end{equation}
    where
    \begin{equation}
        \mathbf{\Phi}_{k}=\sum_{j \neq k}^{K}\left(\rho \mathbf{\lambda}_{k, j}+\rho \mathbf{U}_{k}^{H} \mathbf{H}_{k} \mathbf{V}_{j}\right)\left(\rho \mathbf{\lambda}_{k, j}^{H}+\rho \mathbf{V}_{j}^{H} \mathbf{H}_{k}^{H} \mathbf{U}_{k}\right),
    \end{equation} 
    and 
    \begin{equation}
        \begin{aligned}
            \mathbf{G}_{k}=
            &\rho^{2} \mathbf{D}_{k}^{H}\left(\mathbf{\lambda}_{k, k}+\mathbf{U}_{k}^{H} \mathbf{H}_{k} \mathbf{V}_{k}-\mathbf{I}\right) \\
            &\left(\mathbf{\lambda}_{k, k}^{H}+\mathbf{V}_{k}^{H} \mathbf{H}_{k}^{H} \mathbf{U}_{k}-\mathbf{I}\right) \mathbf{D}_{k}.
        \end{aligned}
    \end{equation}
    It can be seen that the left hand of \eqref{equation_tau} is a decreasing function with respect to $\tau_{k}$, thus, the optimal $\tau_{k}^{*}$ can be attained by the bisection method.
\end{appendices}

\vspace{10pt}

	
\end{document}